\documentclass[10pt]{JHEP3}
\usepackage{epsfig,multicol,bbm}
\usepackage{epsfig}
\usepackage{amsmath}
\usepackage{epsfig}
\usepackage{psfrag}
\usepackage{amsfonts}
\usepackage{graphicx}
\usepackage{dcolumn}
\usepackage{bm}
\usepackage{lineno}
\newcommand{\vect}{\left ( \begin{array}{c}}
\newcommand{\evect}{\end{array} \right )}

\newcommand{\bea}{\begin{eqnarray}}
\newcommand{\eea}{\end{eqnarray}}
\usepackage{epsfig}
\usepackage{psfrag}
\usepackage{graphicx}
\usepackage{dcolumn}
\usepackage{bm}
\def\fsl#1{\setbox0=\hbox{$#1$}                 
   \dimen0=\wd0                                 
   \setbox1=\hbox{/} \dimen1=\wd1               
   \ifdim\dimen0>\dimen1                        
      \rlap{\hbox to \dimen0{\hfil/\hfil}}      
      #1                                        
   \else                                        
      \rlap{\hbox to \dimen1{\hfil$#1$\hfil}}   
      /                                         
   \fi}                                         %

\title{Some Field Theoretic Issues Regarding the Chiral Magnetic Effect}

\author{De-fu Hou$^{\dagger a}$,
Hui Liu $^{\dagger b}$ and Hai-cang Ren $^{\dagger c}$,$^{\dagger a}$\\

{$^{\dagger a}$ Institute of Particle Physics, Huazhong Normal
University, Wuhan 430079, China}\\
{E-mail:~hdf@iopp.ccnu.edu.cn}\\
{$^{\dagger b}$Physics Department, Jinan University, Guangzhou, China}\\
{E-mail:~tliuhui@jnu.edu.cn}\\
{$^{\dagger c}$Physics Department, The Rockefeller University,
1230 York Avenue, New York, NY 10021-6399} \\
{E-mail:~ren@mail.rockefeller.edu}\\
 }

\abstract {In this paper, we shall address some field theoretic issues regarding the chiral magnetic effect. The general structure of the
chiral magnetic current consistent with the electromagnetic gauge invariance is obtained and the impact of the infrared divergence is examined. Some subtleties on the relation between the chiral magnetic effect and the axial
anomaly are clarified through a careful examination of the infrared limit of the relevant thermal diagrams.}

\begin{document}


\section{Introduction}

The chiral magnetic effect (CME) proposed in \cite{kharzeev,kz,Kharzeev:2007jp,fukushima} provides a new probe of the QCD phase transition and the formation of quark-gluon plasma via relativistic heavy ion collisions(RHIC). The physical picture of CME relies on the interplay between the helicity of a quark and the external magnetic field. Consider a quark with positive (negative) helicity, its magnetic moment and the electric current it carries are always parallel (antiparallel) independently of the sign of its electric charge. The magnetic moment tends to be parallel to the magnetic field, so the electric current will be parallel (antiparallel) to the field for positive (negative) helicity. For massless quarks, the helicity coincides with the axial charge,
\begin{equation}
Q_5=\int d^3\mathbf{r}\bar\psi\gamma_4\gamma_5\psi
\label{naive}
\end{equation}
with the quark spinor $\psi$ carrying both color and flavor indexes.
Therefore, for QGP of a nonzero axial charge density, a net electric current will be generated in (opposite to) the direction of the external magnetic field if the positive (negative) helicity is in excess.

The conditions that support CME are likely implemented in RHIC. Firstly, for off-central collisions, a strong magnetic field is produced perpendicular to the collision plane; Secondly, because of the high temperature, there may be a sizable probability for the transition to a topologically nontrivial gluon configuration accompanied by a change of the axial charge according to the
winding number \cite{Kharzeev:2007jp,sphaleron, gdmoore}
\begin{equation}
\Delta Q_5=n_W\equiv-\frac{N_fg^2}{32\pi^2}\int d^4x\epsilon_{\mu\nu\rho\lambda}F_{\mu\nu}^lF_{\rho\lambda}^l,
\label{deltaq5}
\end{equation}
where $F_{\mu\nu}^l$ is the strength of the color $SU(N_c)$ field ($N_c=3$) with $l$ the color index and $N_f$ is the number of flavors. Thirdly, the de-confined quarks that carry the chiral magnetic current
can travel sufficiently far before hadronization to lead to observable charge asymmetry perpendicular to the collision plane. It has been suggested recently that such a charge asymmetry is correlated with the baryon number asymmetry through a similar mechanism,
the chiral vortical effect \cite{khdt, Son}.
For the experimental status of CME, see for example \cite{star08, star09, Wang09, Asakawa:2010bu}

The chiral magnetic effect for a free quark gas in a static and homogeneous magnetic field
$\mathbf{\cal B}$ at thermal equilibrium has been analyzed in great details. With the aid of the grand partition function at a nonzero axial chemical potential $\mu_5$,
\begin{equation}
Z={\rm Tr}e^{-\beta(H-\mu N-\mu_5Q_5)}
\label{grand}
\end{equation}
with $H$ the Hamiltonian, $N$ the quark number, $\beta$ the inverse temperature and $\mu$ the quark number chemical potential, one obtains the chiral magnetic current $\mathbf{J}=\eta\mathbf{j}$ where
\begin{equation}
\eta=N_c\sum_fq_f^2
\end{equation}
with $q_f$ the charge number of the flavor $f$ and
the current per unit charge given by the classical expression
\begin{equation}
\mathbf{j}=\frac{e^2}{2\pi^2}\mu_5\mathbf{{\cal B}}.
\label{classical}
\end{equation}
The chiral magnetic current at nonzero momentum and frequency has also been calculated via current-current correlator to one loop order within the same grand canonical ensemble defined by (\ref{grand}) \cite{kw}.  The same effect has also been examined with holographic models
\cite{HUYee,rebhan,gorsky,rubakov,gynther,brits,kirsch} and the lattice simulation \cite{Buividovich:2009wi,Buividovich:2010tn}.
The effect of a nonzero quark mass has been considered recently in \cite{wjfu}. A diagrammatic proof of (\ref{classical}) to
all orders at high density has been attempted in \cite{hong}.

It was pointed out in \cite{rubakov} that the naive axial charge (\ref{naive}) is not the right object to define the grand canonical ensemble since it is not conserved because of the axial anomaly,
\begin{equation}
\frac{\partial J_{5\mu}}{\partial x_\mu}
=i\frac{N_fg^2}{32\pi^2}\epsilon_{\mu\nu\rho\lambda}F_{\mu\nu}^lF_{\rho\lambda}^l
+i\eta\frac{e^2}{16\pi^2}\epsilon_{\mu\nu\rho\lambda}F_{\mu\nu}F_{\rho\lambda}
=\frac{\partial\Omega_\mu}{\partial x_\mu},
\label{anomaly0}
\end{equation}
where the axial vector current $J_{5\mu}=i\bar\psi\gamma_\mu\gamma_5\psi$ and
$\Omega_\mu$ is a linear combination of the Chern-Simons of QCD and QED, given by
\begin{equation}
\Omega_\mu=i\frac{N_fg^2}{8\pi^2}\epsilon_{\mu\nu\rho\lambda}
A_\nu^l\left(\frac{\partial A_\lambda^l}{\partial x_\rho}-\frac{1}{3}f^{lab}A_\rho^aA_\lambda^b\right)
+i\eta\frac{e^2}{4\pi^2}\epsilon_{\mu\nu\rho\lambda}A_\nu\frac{\partial A_\lambda}{\partial x_\rho}.
\end{equation}
with $A_\mu^l$ and $A_\mu$ the gauge potential of gluons and photons.
The integration of (\ref{anomaly0}) gives rise to (\ref{deltaq5}), to which the trivial topology of the electromagnetic field does not contribute.
The conserved axial charge to replace $Q_5$ in (\ref{grand}) reads
\begin{equation}
\label{conserved}
\tilde Q_5 = Q_5+i\int d^3\mathbf{r}\Omega_4.
\end{equation}
In what follows, we shall name $Q_5$ the naive axial charge.
Furthermore, the author of \cite{rubakov} argued that the gauge invariance prevents a nonzero chiral magnetic current to be generated from the grand canonical ensemble defined with $Q_5$ and the chiral magnetic current comes
solely from the second term of (\ref{conserved}) in the ensemble defined by $\tilde Q_5$. Because this
term stems from the anomaly, which is universal to all orders, the classical expression
(\ref{classical}) is robust against higher order corrections.

In this paper, we shall analyze the chiral magnetic effect via the current-current correlator in the light of Ref. \cite{rubakov}.
There are standard recipes to implement gauge invariant regularization schemes via thermal diagrams employed in this work. Higher order corrections
can also be included systematically. We find that the validity of the statement in \cite{rubakov} relies on the existence of the infrared limits of the energies and momenta involved, which is not always guaranteed. We shall pinpoint a few exceptions to the statement in \cite{rubakov}, one is caused by the massless poles of the invariant form factors underlying the triangle diagram at $T=0$ and $\mu=0$  and others are related to the noncommutativity between the zero momentum limit and the zero energy limit at $T\ne 0$ and/or $\mu\ne 0$. The latter subtlety is a common feature of thermal field theories. The difference between different orders of limits is likely to be subject to higher order corrections.
Since the magnetic field in RHIC is neither homogeneous nor static and the system is not in a complete thermal equilibrium, these
issues need to be addressed to assess the robustness of the effect in RHIC phenomenology.

In the next section, we shall work out the most general structure of the chiral magnetic current consistent with the rotation symmetry, Bose symmetry and the gauge invariance.
We shall restrict our attention to the diagrams that contribute to the same powers of $\mu_5$ and $\mathbf{{\cal B}}$ as (\ref{classical}).
The infrared subtlety is allocated to some invariant form factors of three point functions. The one-loop evaluation of the chiral magnetic current will be revisited in the section III with the Pauli-Villars regularization. In the section IV, we shall clarify the relation between the chiral magnetic current and the axial anomaly for an inhomogeneous and time-dependent $\mu_5$, which is related to the QGP off thermal equilibrium. The section V will conclude the paper with some open questions.

Throughout the paper, all four momenta will be denoted by capital letters. We shall adopt the Euclidean metric $(1,1,1,1)$ in which a Minkowski four momentum $P=(\mathbf{p},ip_0)$ with  $p_0$ real. All gamma matrices are hermitian.

\section{The General Structure of the Chiral Magnetic Current}


The Lagrange density of a quark matter at nonzero baryon number and axial charge densities
is given by
\begin{eqnarray}
{\cal L} &=& -\frac{1}{4}F_{\mu\nu}^lF_{\mu\nu}^l-\frac{1}{4}F_{\mu\nu}F_{\mu\nu}
-\bar\psi\left(\gamma_\mu\frac{\partial}{\partial x_\mu}-igT^lA_\mu^l-ie\hat q A_\mu\right)\psi
\\\nonumber
&+& \mu\bar\psi\gamma_4\psi+\mu_5\left(\bar\psi\gamma_4\gamma_5\psi
+i\Omega_4\right)+J_\mu^{\rm ext.} A_\mu\\\nonumber
&+& \hbox{gauge fixing terms and renormalization counter terms}
\label{lagrange}
\end{eqnarray}
where $\hat q$ is the diagonal matrix of electric charge in flavor space, $\mu$ is the quark number chemical potential
and $\mu_5$ is the axial charge chemical potential. An external electric current $J_\mu^{\rm ext.}$ has been added to the Lagrange.

The generating functional of the connected Green function of photons is the logarithm
of the partition function
\begin{equation}
Z[J^{\rm ext.}]=\int[dA^l][dA][d\psi][d\bar\psi]\exp\left(i\int dt d^3\mathbf{r}{\cal L}\right).
\end{equation}
For the Matsubara Green functions, the time integral inside $exp(...)$ is along the imaginary
axis of the complex $t$-plane extending from 0 to $i\beta=i/T$ subject to periodic (antiperiodic)
boundary conditions for bosonic (fermionic) field variables and $-T\ln Z$ is the thermodynamic potential
at equilibrium. For the closed time path Green function
(CTP), the time $t$ is integrated along the real axis from $-\infty$ to $\infty$ and then
from $\infty$ back to $-\infty$ and the thermal equilibrium is implemented by the initial
correlations. All
fields can take values on either branch of this contour, which
doubles the number of degrees of freedom \cite{kcchou,keld,schw, rep-145,PeterH}. See appendix \ref{CTP_appendix} for a brief introduction
of the CTP formalism. The external current $J_\mu^{\rm ext.}$
generates a nonzero thermal average of the electromagnetic potential, given by
\begin{equation}
{\cal A}_\mu(x) =-i\frac{\delta \ln Z}{\delta J_\mu^{\rm ext.}(x)}
\end{equation}
and its Legrendre transformation reads
\begin{equation}
\frac{\delta{\cal S}}{\delta{\cal A}_\mu(x)}=-J_\mu^{\rm ext.}(x),
\label{legendre}
\end{equation}
where the effective action
\begin{eqnarray}
\label{effective}
{\cal S}[{\cal A}] &=& -i\ln Z[J^{\rm ext.}]-\int d^4xJ_\mu^{\rm ext.}{\cal A}_\mu\\\nonumber
&=& \int d^4x\left(-\frac{1}{4}{\cal F}_{\mu\nu}{\cal F}_{\mu\nu}
+\eta\frac{e^2}{4\pi^2}\mu_5{\cal A}_i{\cal B}_i\right)
+\Gamma[{\cal A}],
\end{eqnarray}
with ${\cal F}_{\mu\nu}=\frac{\partial {\cal A}_\nu}{\partial x_\mu}
-\frac{\partial {\cal A}_\mu}{\partial x_\nu}$ and
${\cal B}_i=\frac{1}{2}\epsilon_{ijk}{\cal F}_{jk}=(\vec\nabla\times\mathbf{{\cal A}})_i$.
In the second line of (\ref{effective}),
we have separated the contributions of tree diagrams (first two terms) from that of loop diagrams (third term). Eq.(\ref{legendre}) is equivalent
to the Maxwell equation
\begin{equation}
\frac{\partial{\cal F_{\mu\nu}}}{\partial x_\nu}=J_\mu^{\rm ext.}+J_\mu,
\end{equation}
where
\begin{equation}
J_i(x)=\frac{\delta\Gamma}{\delta{\cal A}_i(x)}+\eta\frac{e^2}{2\pi^2}\mu_5{\cal B}_i
\label{extJ}
\end{equation}
represents the induced current in the medium. The functional $\Gamma[{\cal A}]$ can be expanded according to the powers of ${\cal A}$
with the proper vertex functions as coefficients. We have, in momentum space
\begin{equation}
\Gamma[{\cal A}]=\int\frac{d^4Q}{(2\pi)^4}\Big[
-\frac{1}{2}\Pi_{\mu\nu}(Q){\cal A}_\mu^*(Q){\cal A}_\nu(Q)+O({\cal A}^3)\Big],
\label{GammaA}
\end{equation}
where only the term contributing to the linear response is displayed explicitly.
It follows from (\ref{extJ}) that
\begin{equation}
J_i(Q)={\cal K}_{ij}(Q){\cal A}_j(Q),
\label{curr}
\end{equation}
where
\begin{equation}
{\cal K}_{ij}(Q)=-\Pi_{ij}(Q)-i\eta\frac{e^2}{4\pi^2}\mu_5\epsilon_{ijk}q_k
+O({\cal A}^2)
\end{equation}
with all QCD and higher order QED corrections contained in the photon self-energy tensor $\Pi_{\mu\nu}(Q)$. The prescription
of the functional derivative for the retarded linear response is outlined near the end of the appendix \ref{CTP_appendix}.
The antisymmetric part of ${\cal K}_{ij}(Q)$,
\begin{equation}
{\cal K}_{ij}^A(Q)\equiv\frac{1}{2}[{\cal K}_{ij}(Q)-{\cal K}_{ji}(Q)]
\end{equation}
which is odd in $\mu_5$, carries odd parity and generates the chiral magnetic current.

\begin{figure}
\begin{center}
  \includegraphics[width=0.3\linewidth]{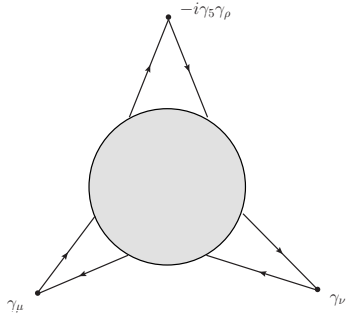}\\
  \caption{The diagrammatic representation of the contribution to the chiral magnetic current from the photon self-energy, where the contribution of each vertex to the Feynman amplitude is indicated explicitly.}\label{fig1}
  \end{center}
\end{figure}

\begin{figure}
\begin{center}
  \includegraphics[width=0.6\linewidth]{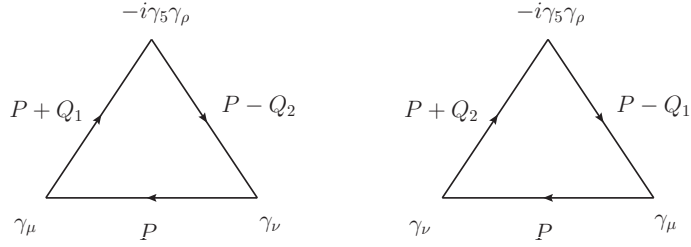}\\
  \caption{The triangle diagram underlying the axial anomaly, where the solid line represents the free quark propagator at $\mu_5=0$}\label{fig2}
  \end{center}
\end{figure}

Expanding the response function ${\cal K}_{ij}^A(Q)$ in the powers of $\mu_5$, we have
${\cal K}_{ij}^A(Q)=\mu_5{\cal K}_{ij}^{(1)}(Q)+O(\mu_5^3)$, where
\begin{equation}
{\cal K}_{ij}^{(1)}(Q)=-\frac{\partial}{\partial\mu_5}\Pi_{\mu\nu}(Q)|_{\mu_5=0}
-i\eta\frac{e^2}{2\pi^2}\epsilon_{ijk}q_k
\label{K1}
\end{equation}
and underlies the classical form of the chiral magnetic current (\ref{classical}).

The first term of (\ref{K1}) is represented
by the 1PI diagram with two external vector vertices and an external axial vector vertex,
shown in Fig.1, at $\mu=i$, $\nu=j$ and $\rho=4$. The lowest order of it consists of the usual triangle diagrams in Fig.2. Let the incoming 4-momenta at the photon
vertices be $Q_1\equiv(\mathbf{q}_1,i\omega)$ and $Q_2\equiv(\mathbf{q_2},-i\omega)$, the incoming 4-momentum at
the axial vertex is then $-Q_1-Q_2=(-\mathbf{q}_1-\mathbf{q}_2,0)$.
The amplitude of the diagram $\Delta_{\mu\nu}(Q_1,Q_2)$ consists of a pseudo-tensor $\Delta_{ij}(Q_1,Q_2)$,
a pseudo-vector $\Delta_{4j}(Q_1,Q_2)$ and a pseudo-scalar $\Delta_{44}(Q_1,Q_2)$. In the limit
$Q_1\to -Q_2$ with $Q_1\equiv Q=(\mathbf{q},i\omega)$, we find that
\begin{equation}
\frac{\partial}{\partial\mu_5}\Pi_{\mu\nu}(Q)|_{\mu_5=0}= \Delta_{\mu\nu}(Q,-Q).
\label{Delta}
\end{equation}
The rotation invariance and the Bose symmetry
\begin{equation}
\Delta_{\mu\nu}(Q_1,Q_2)=\Delta_{\nu\mu}(Q_2,Q_1)
\end{equation}
dictates the following most general tensorial structure
\begin{eqnarray}
\label{tensor}
\Delta_{ij}(Q_1,Q_2) &=& i\eta\frac{e^2}{2\pi^2}
		[C_0(q_1^2,q_2^2,\mathbf{q}_1\cdot\mathbf{q}_2;\omega)\epsilon_{ijk}q_{1k}
                      -C_0(q_2^2,q_1^2,\mathbf{q}_1\cdot\mathbf{q}_2;-\omega)\epsilon_{ijk}q_{2k}\\
           \nonumber &+& C_1(q_1^2,q_2^2,\mathbf{q}_1\cdot\mathbf{q}_2;\omega)\epsilon_{jkl}q_{1k}q_{2l}q_{1i}
                -C_1(q_2^2,q_1^2,\mathbf{q}_1\cdot\mathbf{q}_2;-\omega)\epsilon_{ikl}q_{1k}q_{2l}q_{2j}],
\end{eqnarray}
\begin{equation}
\Delta_{4k}(Q_1,Q_2)= \eta\frac{e^2}{2\pi^2}
	C_2(q_1^2,q_2^2,\mathbf{q}_1\cdot\mathbf{q}_2;\omega)\epsilon_{ijk}q_{1i}q_{2j}
	=\Delta_{k4}(Q_2,Q_1)
\label{vector}
\end{equation}
and $\Delta_{44}(Q_1,Q_2)=0$,
where $C_0$, $C_1$ and $C_2$ are dynamical form factors. The time reversal invariance implies that $C_0$, $C_1$
are even functions of $\omega$ and $C_2$ is odd in $\omega$ (This, however, is not required for our purpose). Notice that the tensors $\epsilon_{ikl}q_{1k}q_{2l}q_{2i}$
and $\epsilon_{ikl}q_{1k}q_{2l}q_{1j}$ are not independent and can be reduced to the tensors already included
in (\ref{tensor}) via Schouten identity
\begin{equation}
\epsilon_{ijk}q_l-\epsilon_{lij}q_k+\epsilon_{kli}q_j-\epsilon_{jkl}q_i=0.
\end{equation}
Furthermore, switching $Q_1\to Q_2$ amounts to $\mathbf{q}_1\to \mathbf{q}_2$ and $\omega\to-\omega$. It follows from (\ref{K1}) and (\ref{tensor}) that
\begin{equation}
{\cal K}_{ij}^A(Q)=i\eta\frac{e^2}{2\pi^2}\mu_5[F(Q)-1]\epsilon_{ijk}q_k+O(\mu_5^3)
\end{equation}
with
\begin{equation}
F(Q)=-C_0(q^2,q^2,-q^2;\omega)-C_0(q^2,q^2,-q^2;-\omega).
\label{FinC0}
\end{equation}
The chiral magnetic current in a constant magnetic field corresponds to the limit
$F(0)$, which is subtle as we shall see.

The electromagnetic gauge invariance,
\begin{equation}
Q_{1\mu}\Delta_{\mu\nu}(Q_1,Q_2)=Q_{2\nu}\Delta_{\mu\nu}(Q_1,Q_2)=0
\end{equation}
gives rise to the relations
\begin{equation}
C_0(q_1^2,q_2^2,\mathbf{q}_1\cdot\mathbf{q}_2;\omega)= -q_2^2C_1(q_2^2,q_1^2,\mathbf{q}_1\cdot\mathbf{q}_2;-\omega)
+\omega C_2(q_2^2,q_1^2,\mathbf{q}_1\cdot\mathbf{q}_2;-\omega)
\end{equation}
and
\begin{equation}
C_0(q_2^2,q_1^2,\mathbf{q}_1\cdot\mathbf{q}_2;-\omega)= -q_1^2C_1(q_1^2,q_2^2,\mathbf{q}_1\cdot\mathbf{q}_2;\omega)
-\omega C_2(q_1^2,q_2^2,\mathbf{q}_1\cdot\mathbf{q}_2;\omega).
\end{equation}
and therefore
\begin{equation}
F(Q)=q^2[C_1(q^2,q^2,-q^2;\omega)+C_1(q^2,q^2,-q^2;-\omega)]
+\omega[C_2(q^2,q^2,-q^2;\omega)-C_2(q^2,q^2,-q^2;-\omega)].
\end{equation}

If the infrared limit of the dynamical form factors $C_1$ and $C_2$ exists, then $F(0)$=0 and there is no
chiral magnetic current associated to the {\it{naive}} axial charge. This is the case in the static limit $q\to 0$ with $Q=(\mathbf{q},0)$ to one-loop order at nonzero $T$ and/or $\mu$. It
remains so if there exists an nonperturbative IR cutoff to remove the $\frac{1}{q^2}$
singularities brought about by QCD corrections\cite{linde} (Such kind of singularities is likely to occur for diagrams with more than one quark loops linked by gluon lines). In that case, the chiral magnetic current takes the classical form (\ref{classical}) to all orders.

It is a common feature of thermal field theories that the  different orders of the double limits $\lim_{q\to 0}\lim_{\omega\to 0}$ and $\lim_{\omega\to 0}\lim_{q\to 0}$ may not agree. While the former order of limits of $C_1(q^2,q^2,-q^2,\omega)$ and $C_2(q^2,q^2,-q^2;\omega)$ converges and leads to the classical form of the chiral magnetic current, the latter order of limits leads
to IR divergence. The explicit calculation of the triangle diagram of Fig.2 in the appendix \ref{IR_formfactor_appendix} with $\mu=4$, $\rho=4$ and $\nu=j$ yields
\begin{equation}
C_2(0,0,0;\omega)=\frac{1}{3\omega}
\label{infrared}
\end{equation}
as $\omega\to 0$ and $\lim_{\omega\to 0}\lim_{q\to 0}F(Q)=\frac{2}{3}$. Consequently, the magnitude of the one-loop chiral magnetic current is reduced to one third of the classical magnitude. This is consistent with the direct one-loop calculation in the literature
\cite{kw} and will be reexamined in the next section. Since the form factor $F(Q)$ is not
linked to the axial anomaly, the chiral magnetic current in this order of limits is likely to be subject to higher order corrections.

The IR singularity also shows up via the massless poles if the zero temperature
and zero chemical potential limits are taken prior to the limit $Q\to 0$ and $\Delta_{\mu\nu}(Q_1,Q_2)$ becomes
fully covariant then. To the one-loop order, the triangle diagram Fig.2 gives rise to
\begin{equation}
C_1(q^2,q^2,-q^2;\omega)=\frac{1}{2(q^2-\omega^2)}
\label{IRc1}
\end{equation}
and
\begin{equation}
C_2(q^2,q^2,-q^2;\omega)=-\frac{\omega}{2(q^2-\omega^2)}.
\label{IRc2}
\end{equation}
(See section IV for details.) Both $C_1$ and $C_2$ are infrared divergent and we find
$F(0)=1$ and therefore zero chiral magnetic current for $T=\mu=0$ but $\mu_5\ne 0$.


\section{The one-loop contribution}

\begin{figure}
\begin{center}
  \includegraphics[width=0.3\linewidth]{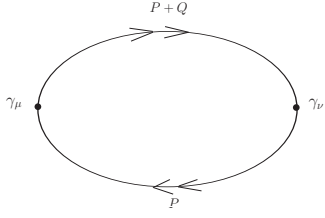}\\
  \caption{The one-loop diagram of the photon self-energy. The solid line with a double arrow stands for the free propagator to all orders of $\mu_5$}\label{fig3}
  \end{center}
\end{figure}

The one-loop contribution to the chiral magnetic current has been discussed extensively in
the literature. In the present section, we shall supplement this calculation with the Pauli-Villars regularization, since the photon self-energy as a whole suffers from the
UV divergence. As the regularization respects the gauge invariance, the result will be
consistent with the Ref.\cite{rubakov} and the statement of the previous section. The trivial color-flavor factor $\eta$ will be suppressed below.

The one-loop photon self-energy tensor at the temperature $T$, shown in Fig.3 is given by
\begin{equation}
\Pi_{\mu\nu}(Q)=e^2T\sum_{p_0}\int\frac{d^3\mathbf{p}}{(2\pi)^3}
\Big[\Xi_{\mu\nu}(P,Q|m)-\sum_sC_s\Xi_{\mu\nu}(P,Q|M_s)\Big],
\end{equation}
where
\begin{equation}
\Xi_{\mu\nu}(P,Q|m)={\rm tr}S_F(P+Q|m)\gamma_\mu S_F(P|m)\gamma_\nu.
\label{Xi}
\end{equation}
and the summation in the integrand corresponds to the contribution of the Pauli-Villars regulators
that remove all UV divergences. We have
\begin{equation}
\sum_sC_s=1
\label{PV}
\end{equation}
and $M_s\longrightarrow\infty$ after the integration.
The free quark propagator with a four momentum $P=(\mathbf{p},ip_0)$, a mass $m$, a quark number chemical potential $\mu$ and an axial charge chemical potential $\mu_5$ reads
\begin{eqnarray}
S_F(P|m)&=&\frac{i}{{\not P}+\mu\gamma_4+\mu_5\gamma_4\gamma_5-m}\\\nonumber
&=&\frac{i}{2}[A(P,m,\mu,\mu_5)+A(P,m,\mu,-\mu_5)]
+\frac{i}{2}\gamma_5[B(P,m,\mu,\mu_5)-B(P,m,\mu,-\mu_5)]
\label{prop}
\end{eqnarray}
where ${\not P}\equiv \gamma_4 p_0-i\mathbf{\gamma}\cdot\mathbf{p}$ and
we have decomposed $S_F(P|m)$ into the parts even and odd in $\mu_5$ with
\begin{equation}
A(P,m,\mu,\mu_5)=\frac{(p_0+\mu)\gamma_4-i(p+\mu_5)\mathbf{\gamma}\cdot
\hat{\mathbf{p}}+m}{(p_0+\mu)^2-(p+\mu_5)^2-m^2}
\label{AP}
\end{equation}
and
\begin{equation}
B(P,m,\mu,\mu_5)=\frac{-(p+\mu_5)\gamma_4+i(p_0+\mu-m\gamma_4)\mathbf{\gamma}
\cdot\hat{\mathbf{p}}}{(p_0+\mu)^2-(p+\mu_5)^2-m^2}.
\label{BP}
\end{equation}
The chiral magnetic current corresponds to the antisymmetric spatial components
of $\Pi_{\mu\nu}(Q)$, i.e.
\begin{equation}
\Pi_{ij}^A(Q)\equiv\frac{1}{2}[\Pi_{ij}(Q)-\Pi_{ji}(Q)]=
e^2T\sum_{p_0}\int\frac{d^3\mathbf{p}}{(2\pi)^3}
\Big[\Xi_{ij}^A(P,Q|m)-\sum_sC_s\Xi_{ij}^A(P,Q|M_s)\Big],
\end{equation}
where
\begin{eqnarray}\label{XiA}
\Xi_{ij}^A(P,Q|m)&=&-\frac{1}{4}{\rm tr}\gamma_5
\lbrace[B(P+Q,m,\mu,\mu_5)-B(P+Q,m,\mu,-\mu_5)]\gamma_i\\\nonumber
&\times& [A(P,m,\mu,\mu_5)+A(P,m,\mu,-\mu_5)]\gamma_j\\\nonumber
&+&[B(P,m,\mu,\mu_5)-B(P,m,\mu,-\mu_5)]\gamma_i
[A(P+Q,m,\mu,\mu_5)+A(P+Q,m,\mu,-\mu_5)]\gamma_j\rbrace.
\end{eqnarray}
It is straightforward to work out the trace and summation over  the  Matsubara frequency,
$p_0=i(2n+1)\pi T$. To obtain the retarded self-energy, we shall follow the recipe of Baym and Mermin \cite{baym}to extend the Matsubara frequency $q_0$ to the upper edge of the real axis, $q_0\to \omega+i0^+$. The details are shown in the appendix \ref{oneloop_appendix} and we shall report two special cases below. The antisymmetric part of the self-energy tensor is parametrized as
\begin{equation}
\Pi_{ij}^A(Q)=-i\frac{e^2}{2\pi^2}\mu_5F_1(q,\omega)\epsilon_{ijk}q_k,
\label{form}
\end{equation}
with $F_1(q,\omega)$ at $\mu_5=0$ corresponds to the one-loop approximation of $F(q,\omega)$ as defined in Eq. (\ref{FinC0}).
The dependences on
the spatial momentum and the energy are indicated separately here. Diagrammatically, Fig.2 corresponds to the linear term of
the Taylor expansion of Fig.3 in $\mu_5$.

\noindent
\subsection{The static limit}

At zero frequency, $q_0=0$, we find that
\begin{equation}
F_1(q,0)=-{\cal F}(q|m)+\sum_sC_s{\cal F}(q|M_s)
\label{static}
\end{equation}
where

\begin{eqnarray}
\label{calF}
{\cal F}(q|m)&=&\frac{1}{2\mu_5q}\int_0^\infty dpp\ln|\frac{2p-q}{2p+q}|\\\nonumber
    &{}&\lbrace\frac{p+\mu_5}{E_+}[f(E_+-\mu)-f(-E_+-\mu)]
    -\frac{p-\mu_5}{E_-}[f(E_--\mu)-f(-E_--\mu)]\rbrace
\end{eqnarray}
with
\begin{equation}
E_\pm=\sqrt{(p\pm\mu_5)^2+m^2},
\end{equation}
and the Fermi distribution function
\begin{equation}
f(\xi)=\frac{1}{e^{\beta\xi}+1}.
\label{distr}
\end{equation}
It is straightforward to verify that the limit $q\to 0$ at $T\neq 0$ and/or $\mu\neq 0$ yields,
\begin{eqnarray}
{\cal F}(0|m)&=&-\frac{1}{2\mu_5}\int_0^\infty dp \lbrace[\frac{p+\mu_5}{E_+}
[f(E_+-\mu)-f(-E_+-\mu)]-\frac{p-\mu_5}{E_-}[f(E_--\mu)-f(-E_--\mu)]\rbrace
\nonumber  \\
&=&\frac{1}{2\mu_5\beta}\Big[\ln(1+e^{-\beta(E_+-\mu)})
-\ln(1+e^{-\beta(E_--\mu)})+\ln(e^{\beta(E_++\mu)}+1)
-\ln(e^{\beta(E_-+\mu)}+1)\Big]_0^\infty\nonumber  \\
&=&\frac{1}{2\mu_5}\lim_{p\to\infty}(E_+-E_-)=1,
\label{homo}
\end{eqnarray}
Then eqs.(\ref{static}) and (\ref{PV}) implies that
\begin{equation}
\lim_{q\to 0}\lim_{\omega\to 0}F_1(q,\omega)=0.
\end{equation}
This result is expected according to the discussion in the last section because the
nonzero Matsubara frequency, $(2n+1)\pi T$ regularizes the infrared behavior of the
quark propagator even in the massless limit. Notice that the regulator contribution
\begin{equation}
\lim_{M_s\to \infty}{\cal F}(q|M_s)=1
\label{pvlimit}
\end{equation}
for all $q$ and this is also the case with a non static $Q$.
If, on the other hand, $T$ and $\mu$ as well as the quark mass are set
to zero first, we find
\begin{equation}
{\cal F}(q|0)=-\frac{1}{\mu_5 q}\int_0^{|\mu_5|} dpp\ln\left|\frac{2p-q}{2p+q}\right|.
\label{covlimit}
\end{equation}
It follows from (\ref{pvlimit}), (\ref{covlimit}) and (\ref{static}) that $F_1(q,0)=-1$ at $\mu_5=0$, in agreement with the covariant result reported at the end of the last section.

\subsection{Massless limit}

In the massless limit, $m=0$, the quark propagator (\ref{prop}) reduces to
\begin{equation}
S_F(P|0)=\frac{i}{{\not P}+(\mu-\mu_5)}\frac{1+\gamma_5}{2}
+\frac{i}{{\not P}+(\mu+\mu_5)}\frac{1-\gamma_5}{2}
\label{propm0}
\end{equation}
and $F_1(q,\omega)$ in this case reads
\begin{eqnarray}
\label{massless}
F_1(q,\omega) &=& -\frac{1}{2\mu_5}\int_0^\infty dp p^2\Big[
\frac{J(p,q,\omega)+J(p,-q,-\omega)}{e^{\beta(p-\mu+\mu_5)}+1}
-\frac{J(p,q,-\omega)+J(p,-q,\omega)}{e^{\beta(p+\mu-\mu_5)}+1}\nonumber  \\
&-&\frac{J(p,q,\omega)+J(p,-q,-\omega)}{e^{\beta(p-\mu-\mu_5)}+1}
+\frac{J(p,q,-\omega)+J(p,-q,\omega)}{e^{\beta(p+\mu+\mu_5)}+1}
\Big]+1,
\end{eqnarray}
where
\begin{equation}
{\rm Re}J(p,q,\omega)=\frac{1}{pq}\Big[-\frac{\omega}{q}
+\frac{1}{2}\left(1+\omega\frac{\omega^2-2p\omega-q^2}{2pq^2}\right)
\ln|\frac{(\omega-q)(\omega+q-2p)}{(\omega+q)(\omega-q-2p)}|\Big]
\end{equation}
and
\begin{equation}
{\rm Im}J(p,q,\omega)=\frac{\pi}{pq}{\rm sign}(\omega)
\left(1+\omega\frac{\omega^2-2p\omega-q^2}{2pq^2}\right)
\theta\left(1-\frac{|q^2+2p\omega-\omega^2|}{2pq}\right).
\end{equation}
and the "+1" of eq.(\ref{massless}) comes from the Pauli-Villars
regulators. The limit $Q\to 0$ of the PV regulators is independent of the order between $q\to 0$ and $\omega\to 0$ as long as $M_s\to\infty$ is taken first. The same limit of massless
part is, however, subtle. We have
\begin{equation}
\lim_{q\to 0}\lim_{\omega\to 0}F_1(q,\omega)=0
\end{equation}
but
\begin{equation}
\lim_{\omega\to 0}\lim_{q\to 0}F_1(q,\omega)=\frac{2}{3}
\end{equation}
consistent with the result reported in \cite{kw}. The nonzero value of
the latter limit signals infrared divergence of the form factor
$C_2(q^2,q^2,-q^2;\omega)$ defined in the last
section under the same orders of limits.


\section{The Relation to the Triangle Anomaly}

In the section 2, we related the chiral magnetic current to the infrared limit of the
three point Green's function in Fig.1 with two electric currents and the fourth component
of the axial vector current. We analyzed the general structure of the chiral magnetic
current as is required by the electromagnetic Ward identity.
For the sake of simplicity, we restricted our attention to zero energy flow at the axial
vector vertex. To explore the the impact of the anomalous axial current Ward identity,
this restriction will be relaxed in the present section. The physics of the diagram of Fig. 1
with $\rho=4$ and an arbitrary $Q_1+Q_2$ corresponds to
CME at a space-time dependent $\mu_5$ in a QGP off thermal equilibrium.

We shall denote the general Feynman amplitude of Fig.1 by $\Lambda_{\mu\nu\rho}(Q_1,Q_2)$ with $Q_1$ and $Q_2$ the incoming momenta at the vector
vertices indexed by $\mu$ and $\nu$. We have
\begin{equation}
\Lambda_{\mu\nu 4}(Q_1,Q_2)=\Delta_{\mu\nu}(Q_1,Q_2)
\end{equation}
with $\Delta_{\mu\nu}(Q_1,Q_2)$ defined in the section 2.
The incoming momentum at the axial vector vertex is then
\begin{equation}
K=(\mathbf{k},ik_0)=-Q_1-Q_2.
\label{outgoing}
\end{equation}
We have
\begin{equation}
Q_{1\mu}\Lambda_{\mu\nu\rho}(Q_1,Q_2)=Q_{2\nu}\Lambda_{\mu\nu\rho}(Q_1,Q_2)=0
\label{gaugeinv}
\end{equation}
following from the electromagnetic gauge invariance. The triangle anomaly implies that
\begin{equation}
(Q_1+Q_2)_\rho\Lambda_{\mu\nu\rho}(Q_1,Q_2)
=-i\eta\frac{e^2}{2\pi^2}\epsilon_{\mu\nu\alpha\beta}Q_{1\alpha}Q_{2\beta}
\label{anomaly}
\end{equation}
which holds to all orders of interaction at arbitrary temperature and chemical potential
\cite{itoyama}.
The classical expression of the chiral magnetic current is associated to the
component $\Lambda_{ij4}(Q_1,Q_2)$ with the momenta
\begin{equation}
Q_1=(\mathbf{q},i\omega) \qquad Q_2=(-\mathbf{q},-i\omega).
\label{cmemomenta}
\end{equation}

It is tempting to relate the self-energy contribution to CME with the axial anomaly via the limiting process
\begin{equation}
\Lambda_{ij4}(Q_1,Q_2)=-i\lim_{k_0\to 0}\frac{1}{k_0}
(Q_1^\prime+Q_2^\prime)_\rho\Lambda_{ij\rho}(Q_1^\prime,Q_2^\prime)
=-i\eta\frac{e^2}{2\pi^2}\epsilon_{ijk}q_k
\label{cmelimit}
\end{equation}
where $Q_1^\prime\equiv(\mathbf{q},ik_0/2)$ and $Q_2^\prime\equiv(-\mathbf{q},ik_0/2)$.
This appears in contradiction with the statement of the absence of CME with the naive axial charge.  It  does not  display the nontrivial energy-momentum dependence of the one-loop result. The reason lies in the infrared singularity and the subtlety of the order of limits $k_0\to 0$ and $\mathbf{k}\to 0$
as we shall analyze below. At $T=0$ and $\mu=0$, however, the order of limits is
irrelevant and we always get the RHS of (\ref{cmelimit}), consistent with the one loop result near the end of the subsection 3.1.

The most general tensorial decomposition at $T=\mu=0$ consistent with the
gauge invariance (\ref{gaugeinv}) and Bose symmetry reads
\begin{eqnarray}
\Lambda_{\mu\nu\rho}(Q_1,Q_2) &=& i\eta\frac{e^2}{2\pi^2}\lbrace\epsilon_{\mu\nu\alpha\beta}Q_{1\alpha}Q_{2\beta}
[Q_{1\rho}D_1(Q_1^2,Q_2^2,Q_1\cdot Q_2)+Q_{2\rho}D_1(Q_2^2,Q_1^2,Q_1\cdot Q_2)] \nonumber\\
&+&(\epsilon_{\nu\rho\alpha\beta}Q_{1\alpha}Q_{2\beta}Q_{1\mu}
-Q_1^2\epsilon_{\mu\nu\rho\lambda}Q_{2\lambda})D_2(Q_1^2,Q_2^2,Q_1\cdot Q_2)
\\\nonumber
&-&(\epsilon_{\mu\rho\alpha\beta}Q_{1\alpha}Q_{2\beta}Q_{2\nu}
-Q_2^2\epsilon_{\mu\nu\rho\lambda}Q_{1\lambda})D_2(Q_2^2,Q_1^2,Q_1\cdot Q_2)\rbrace,
\label{*}
\end{eqnarray}
where the 4D Schouten identity
\begin{equation}
\epsilon_{\mu\nu\rho\lambda}Q_\alpha+\epsilon_{\alpha\mu\nu\rho}Q_\lambda
+\epsilon_{\lambda\alpha\mu\nu}Q_\rho+\epsilon_{\rho\lambda\alpha\mu}Q_\nu
+\epsilon_{\nu\rho\lambda\alpha}Q_\mu=0
\end{equation}
is employed to reduce the number of terms.
It follows from the anomaly equation (\ref{anomaly}) that
\begin{eqnarray}
&{}&(Q_1+Q_2)\cdot Q_1D_1(Q_1^2,Q_2^2,Q_1\cdot Q_2)+(Q_1+Q_2)\cdot Q_2D_1(Q_2^2,Q_1^2,Q_1\cdot Q_2) \nonumber\\
&-&Q_1^2D_2(Q_1^2,Q_2^2,Q_1\cdot Q_2)-Q_2^2D_2(Q_2^2,Q_1^2,Q_1\cdot Q_2)=-1,
\label{constraint}
\end{eqnarray}
which implies infrared singularities of the dynamical form factors $D_1$ and $D_2$. To the one loop order,
we find that
\begin{equation}
D_1(Q_1^2,Q_2^2,Q_1\cdot Q_2)=-2\int_0^1dx\int_0^{1-x}dy
\frac{xy}{Q_1^2x+Q_2^2y-(Q_1x-Q_2y)^2}
\end{equation}
and
\begin{equation}
D_2(Q_1^2,Q_2^2,Q_1\cdot Q_2)=2\int_0^1dx\int_0^{1-x}dy
\frac{x(1-x-y)}{Q_1^2x+Q_2^2y-(Q_1x-Q_2y)^2},
\label{D2}
\end{equation}
which satisfy the constraint (\ref{constraint}).
For the CME momenta, (\ref{cmemomenta}), we find that $D_2(Q^2,Q^2,-Q^2)=\frac{1}{2Q^2}$ and therefore
\begin{equation}
\Lambda_{ij4}(Q_1,Q_2)=-i\eta\frac{e^2}{\pi^2}Q^2D_2(Q^2,Q^2,-Q^2)\epsilon_{ijk}q_k
=-i\eta\frac{e^2}{2\pi^2}\epsilon_{ijk}q_k.
\end{equation}
Breaking the tensor (\ref{*}) into spatial and temporal components, we obtain (\ref{IRc1}) and (\ref{IRc2}) via (\ref{D2}).

At a nonzero temperature and/or chemical potential, the limit $K\to 0$ becomes very subtle. Because of the discreteness of the energy in the Matsubara Green's function, one has to switch to the real time formalism for the analysis, of which, the closed time path (CTP) Green's function is most convenient. The main ingredients of CTP is summarized in the appendix \ref{CTP_appendix}. Explicit calculations of the triangle diagram via the CTP show that
\begin{equation}
\lim_{\mathbf{k}\to 0}\lim_{k_0\to 0}\Lambda_{ij4}(Q_1,Q_2)
\neq \lim_{k_0\to 0}\lim_{\mathbf{k}\to 0}\Lambda_{ij4}(Q_1,Q_2).
\label{subtlety}
\end{equation}
with $\mathbf{k}$ and $k_0$ defined in (\ref{outgoing}).
The limit order on RHS leads to (\ref{cmelimit}), the result dictated by the anomaly, while the limit order on LHS gives rise to result of the last section, obtained from the Matsubara formulation and its analytic continuation to real energy. Therefore, there is no contradiction between the universality of the anomaly and the statement of \cite{rubakov}.

\begin{figure}
\begin{center}
  \includegraphics[width=0.2\linewidth]{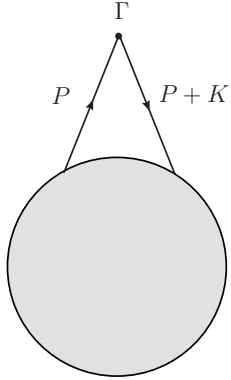}\\
  \caption{The CTP diagram with one vertex insertion highlighted.}\label{fig4}
  \end{center}
\end{figure}

The subtlety of this infrared limit can be explored in general.
Consider the CTP diagram in Fig.4 with a vertex insertion of four momentum $K=(\mathbf{k},ik_0)$, summing up both CTP paths. The amputated
external legs pertaining to the shaded bubble are suppressed.
It follows from the Feynman rules of CTP that the contribution of the two highlighted
lines adjacent to the vertex insertion in the Fig. 4 is
\begin{equation}
S_{1a}(P+K)\Gamma S_{b1}(P)-S_{2a}(P+K)\Gamma S_{b2}(P),
\label{ctp12}
\end{equation}
where $S_{ab}(P)$ is the CTP quark propagator defined in the appendix \ref{CTP_appendix} with $a$, $b$ labeling the two CTP paths and $\Gamma$ is a matrix with respect to the spinor indexes. The spinor indexes as well as the indexes $a$ and $b$ of (\ref{ctp12}) are to be contracted with the contribution from the shaded bubble of Fig. 4. In terms of the retarded(advanced) propagator $S_R(P)$($S_A(P)$) and the correlator $S_C(P)$ defined in Eq.(\ref{3af}), we find that
\begin{eqnarray}
\label{ctpprod}
&{}& S_{1a}(P+K)\Gamma S_{b1}(P)-S_{2a}(P+K)\Gamma S_{b2}(P)\\\nonumber
&=& \frac{1}{2}[S_C(P+K)\Gamma S_R(P)+S_A(P+K)\Gamma S_C(P)
\pm S_R(P+K)\Gamma S_R(P)\pm S_A(P+K)\Gamma S_A(P)].
\end{eqnarray}
with "$\pm$" on RHS depending on the CTP indexes, $a$ and $b$.
Therefore the amplitude of the diagram has the following mathematical structure
\label{generalG}
\begin{eqnarray}
{\cal G}(K)&=&\int\frac{d^4P}{(2\pi)^4}\lbrace U(p_0,\mathbf{p};k_0,\mathbf{k})
\Big[\frac{[1-2f(p^\prime)]\delta[(P+K)^2]}{P^2}+\frac{[1-2f(p)]\delta(P^2)}{(P+K)^2}
\Big]\\\nonumber
&+&V(p_0,\mathbf{p};k_0,\mathbf{k})\rbrace,
\end{eqnarray}
where $p^\prime\equiv|\mathbf{p}+\mathbf{k}|$ and $f(p)$ stands for
the fermion distribution function. For the sake of simplicity, we have set the quark number chemical potential
$\mu=0$, but the generalization to a nonzero $\mu$ is straightforward. The quantity inside the bracket on RHS of
(\ref{generalG}) comes from the first two terms in the second line of eq.(\ref{ctpprod})
and the contribution from the shaded bubble of Fig.4 is included in the functions
$U(p_0,\mathbf{p};k_0,\mathbf{k})$ and $V(p_0,\mathbf{p};k_0,\mathbf{k})$.
The function $U(p_0,\mathbf{p};k_0,\mathbf{k})$ is regular at the mass shells
\begin{equation}
P^2=p^2-p_0^2=0 \qquad (P+K)^2=p^{\prime 2}-p_0^{\prime 2}=0,
\label{massshell}
\end{equation}
and its derivative with respect to $p_0$ will be denoted by
$\dot{U}(p_0,\mathbf{p};k_0,\mathbf{k})$ below. So is the function
$V(p_0,\mathbf{p};k_0,\mathbf{k})$ and its integral,
$I(K)$ is unambiguous in the limit $K\to 0$. Carrying out the
energy integral, we find that
\begin{eqnarray}
{\cal G}(K)=\frac{1}{2}\int\frac{d^3p}{(2\pi)^3} &\lbrace&
\frac{1}{p-p'+k_0}
\Big[\frac{[1-2f(p')]U(p'-k_0,\mathbf{p};k_0,\mathbf{k})}{p'(p+p'-k_0)}
-\frac{[1-2f(p)]U(p,\mathbf{p};k_0,\mathbf{k})}{p(p+p'+k_0)}\Big]\nonumber\\
&+& \frac{1}{p-p'-k_0}
\Big[\frac{[1-2f(p')]U(-p'-k_0,\mathbf{p};k_0,\mathbf{k})}{p'(p+p'+k_0)}\nonumber\\
&-&\frac{[1-2f(p)]U(-p,\mathbf{p};k_0,\mathbf{k})}{p(p+p'-k_0)}\Big]\rbrace
+ I(K).
\end{eqnarray}
It follows that
\begin{eqnarray}
\lim_{k_0\to 0}\lim_{\mathbf{k}\to 0}{\cal G}(K) &=&
\frac{1}{2}\lim_{k_0\to 0}\frac{1}{k_0}\int\frac{d^3p}{(2\pi)^3}
\frac{1-2f(p)}{p}[\frac{U(p-k_0,\mathbf{p};k_0,0)}{(2p-k_0)}
-\frac{U(p,\mathbf{p};k_0,0)}{(2p+k_0)}\nonumber\\
&-&\frac{U(-p-k_0,\mathbf{p};k_0,0)}{(2p+k_0)}+\frac{U(-p,\mathbf{p};k_0,0)}{(2p-k_0)}
+I(0)\Big]\nonumber\\
&=&\frac{1}{4}\int\frac{d^3p}{(2\pi)^3}\frac{1-2f(p)}{p^2}
\Big[-\dot{U}(p,\mathbf{p};0,0)+\frac{U(p,\mathbf{p};0,0)}{p}\nonumber\\
&+& \dot{U}(-p,\mathbf{p};0,0)+\frac{U(-p,\mathbf{p};0,0)}{p}\Big]+I(0)
\end{eqnarray}
and
\begin{eqnarray}
\label{limitorder} \lim_{\mathbf{k}\to 0}\lim_{k_0\to 0}{\cal G}(K) &=& \frac{1}{2}\lim_{\mathbf{k}\to
0}\int\frac{d^3p}{(2\pi)^3}\frac{1}{p^2-p^{\prime 2}}
\lbrace\frac{[1-2f(p')]U(p',\mathbf{p};0,\mathbf{k})}{p'}
       -\frac{[1-2f(p)]U(p,\mathbf{p};0,\mathbf{k})}{p}\nonumber\\
     &+&\frac{[1-2f(p')]U(-p',\mathbf{p};0,\mathbf{k})}{p'}
     -\frac{[1-2f(p)]U(-p,\mathbf{p};0,\mathbf{k})}{p}\rbrace+I(0)\nonumber\\
&=& \lim_{k_0\to 0}\lim_{\mathbf{k}\to 0}{\cal G}(K)
+\frac{1}{2}\int\frac{d^3p}{(2\pi)^3}\frac{df}{dp}\frac{U(p,\mathbf{p};0,0)+U(-p,\mathbf{p};0,0)}{p^2}.
\end{eqnarray}

The inequality (\ref{subtlety}) is an example eq.(\ref{limitorder}) for the three point function $\Lambda_{\mu\nu\rho}(Q_1,Q_2)$ with $\Gamma=-i\gamma_5\gamma_4$. Applying (\ref{limitorder}) for the one loop diagrams of $\Lambda_{ij4}(Q_1,Q_2)$ with
(\ref{cmelimit}) for the first term in the third line, we recover the CME term of the photon self energy obtained previously.
Indeed, the first term in the third line of (\ref{limitorder}) corresponds to the term "+1" on RHS of (\ref{massless}) and the integral of
(\ref{limitorder}) in the same line goes to the integral of (\ref{massless}) in the limit $\mu_5\to 0$. Since the only $\mu_5$ dependence of
(\ref{massless}) is through the distribution functions, the limit has its integrand proportional to the derivative of the distribution function.


\section{Discussions}

In this work, we investigated the interplay between the gauge invariance and the infrared limit in the chiral magnetic effect.
The part of the induced electric current that is
linear in the axial chemical potential $\mu_5$ and the magnetic field $\mathbf{{\cal B}}$ is
divided into two terms, i.e.
\begin{equation}
\mathbf{J}(Q)=-\eta\frac{e^2}{2\pi^2}\mu_5F(Q)\mathbf{{\cal B}}(Q)+\eta\frac{e^2}{2\pi^2}\mu_5\mathbf{{\cal B}}(Q)
\label{division}
\end{equation}
where the first term corresponds to the loop diagrams of the photon self-energy tensor and
the second term comes from the Chern-Simons term of the conserved axial charge $\tilde Q_5$,
which is dictated by the anomaly.
The gauge invariance relates the form factor $F(Q)$ to two form factors,
$C_1$ and $C_2$ underlying a three point diagram of two vector current vertices and an axial current vertex. If the infrared limit of these form factors exists, $F(0)=0$ to all orders of coupling
and the classical form of the chiral magnetic current in a constant magnetic field, eq. (\ref{classical})
emerges. Our statements are illustrated with explicit one-loop calculations subject to the Pauli-Villars regularization. At zero temperature, however, both $C_1$ and $C_2$ are infrared divergent and $F(0)=1$. Consequently,
the two terms on RHS of (\ref{division}) cancel each other and the chiral magnetic current vanishes. At a nonzero temperature and/or a nonzero chemical potential, $F(0)$ depends on how the limit $Q\to 0$
is approached. The magnitude of the chiral magnetic current is reduced if the zero momentum limit is taken prior to the zero energy limit,
as is implied by the infrared divergence of $C_2$ under the same order of limits. More subtle is the situation with a coordinate dependent $\mu_5$.
If the four momentum associated with $\mu_5$, $K=(\mathbf{k},ik_0)$ is set to zero in the order $\lim_{\mathbf{k}\to 0}\lim_{k_0\to 0}$,
the results of sections 2 and 3 are recovered. With the opposite order of the limit, however, $F(0)=1$ as is dictated by the anomaly and the two
terms of (\ref{division}) cancel again. Unlike what happens with the axial anomaly, the difference between different
orders of the infrared limits is unlikely robust against higher order corrections. Since the ambiguity
stems from quasi particle poles, it will disappear when the quasi particle weight is diminished by strong coupling. Then
the chiral magnetic current will revert to its classical expression  with the order $\lim_{\omega\to 0}\lim_{q\to 0}$ of the
infrared limit $Q\to 0$. This is consistent with the holographic result reported in \cite{HUYee}.

One complication with a coordinate dependent $\mu_5$ is that the term $\mu_5\tilde Q_5$ of the Lagrangian (\ref{lagrange})
is no longer gauge invariant. One may argue that this term is only defined in a specific
gauge, say Coulomb gauge, in which the vector potential
\begin{equation}
\mathbf{A}=-\frac{1}{\nabla^2}\mathbf{\nabla}\times\mathbf{B}
\label{coulomb}
\end{equation}
is already gauge invariant. So is the Chern-Simons term of $\tilde Q_5$. A possible objection to this
approach is the violation of the micro causality, i.e. the commutator between two axial charge
densities in Heisenberg representation does not vanish for  a space-like separation because of the nonlocality
introduced by the inverse Laplacian in (\ref{coulomb}). It remains an open issue to assess the validity of
the conserved axial charge in a non equilibrium setup (See \cite{gynther} for some related
discussions).

Finally, we would like to comment briefly on the derivation of the classical result (\ref{classical})
by summing up the single particle Landau orbitals in a constant magnetic field. It is a one-loop procedure
to all orders of the magnetic field. The linear term of the electric and the magnetic field
stems from the same photon self energy tensor discussed here and requires a gauge invariant
regularization to cancel the UV divergence. In view of the analysis in this paper, we would
expect that the summation over the Landau orbitals yields a null result for the chiral magnetic
current if the regulator contribution is included. The net current is solely given by the
Chern-Simons term of the (\ref{lagrange}). Therefore we do not see any nontrivial effect of
a nonzero quark mass  claimed in \cite{wjfu}.

\section*{Acknowledgments}

The work of D. F. H. and H. C. R.
is supported in part by NSFC under grant Nos. 10975060, 10735040. The work of Hui Liu is supported in part by NSFC under grant No. 10947002.



\appendix

\section{Some elements of the closed time path Green functions}
\label{CTP_appendix}
The CTP formalism of a finite temperature field theory was introduced
by Keldysh \cite{keld} and Schwinger \cite{schw}. Good reviews can be
found in \cite{kcchou,rep-145,PeterH}. The CTP contour on the complex time plane has two
branches: ${\cal C}_1$ runs from negative infinity to positive
infinity just above the real axis, and ${\cal C}_2$ runs back from
positive infinity to negative infinity just below the real axis. All
fields can take values on either branch of this contour, which
results in a doubling in the number of degrees of freedom.
 The scalar propagator is given by,
 \bea
 D(X-Y) = \langle
T_c \phi(X) \phi(Y) \rangle
\eea
where $T_c$ is the operator that
time orders along the CTP contour. We also use the notation
$X=(\mathbf{x},it)$ and $P=(\mathbf{p},ip_0)$. The propagator has $2^2=4$
components and can be written as a $2 \times 2$ matrix

 \bea
 \label{2x2}
 D &=&\left(
 \begin{matrix}
    D_{11} & D_{12} \cr
   D_{21} & D_{22}
\end{matrix}\right )
   \nonumber
\eea
with \bea
 D_{11}(X-Y) &=& \langle T(\phi(X) \phi(Y))\rangle \, , \nonumber\\
 D_{12}(X-Y) &=& \langle \phi(Y) \phi(X) \rangle \, , \nonumber\\
 D_{21}(X-Y) &=& \langle \phi(X) \phi(Y)\rangle \, , \nonumber\\
 D_{22}(X-Y) &=& \langle\tilde{T}(\phi(X)\phi(Y))\rangle \, ,\label{D11}
\eea where $T$ is the usual time ordering operator, and $\tilde{T}$
is the anti-chronological time ordering operator. These four
components satisfy,
 \begin{equation}
D_{11} - D_{12} - D_{21} + D_{22} = 0 \nonumber
\end{equation}
as a consequence of the identity $\theta(x) + \theta(-x) =1$.

It is more useful to write the propagator in terms of the three
functions
\begin{eqnarray} \label{3a}
D_R &=& D_{11} - D_{12} \, , \nonumber\\
 D_A &=& D_{11} - D_{21} \, , \nonumber\\
 D_C &=& D_{11} + D_{22} \, .
\end{eqnarray}
$D_R$ and $D_A$ are the usual retarded and advanced propagators,
satisfying
 \begin{equation}
   D_R(X-Y)-D_A(X-Y) = \langle [\phi(X),\phi(Y)] \rangle\, ,\nonumber
 \end{equation}
and $D_C$ is the symmetric combination,also called correlator
 \begin{equation}
   D_C(X-Y) = \langle \{\phi(X),\phi(Y)\} \rangle\, ,\nonumber
 \end{equation}
which satisfies the  Kubo-Martin-Schwinger (KMS) condition at thermal equilibrium. In momentum space \bea
  D_{R,A}(P) &&= \frac{i}{(p_0\pm i\epsilon)^2 - {\vec p}^{\,2}-m^2},\nonumber
\\
D_C(P) &&= (1+2n(p_0))(D_R(P) - D_A(P))
=2\pi[1+2n(E_p)]\delta(P^2+m^2), \label{KMS1} \eea where
$n(p_0)$ is the thermal Bose-Einstein distribution,
 \bea
 \label{Bose}
   n(p_0) = {1 \over e^{\beta p_0} -1}\, ,\qquad
   n(-p_0) = - \bigl( 1 + n(p_0) \bigr)\,
 \eea
and $E_p=\sqrt{p^2+m^2}$. The propagator can be rewritten as\cite{PeterH}:
 \begin{equation}
 \label{decompD1}
   2\,D = D_R {1\choose 1}\left(1,-1\right)
        + D_A {1\choose -1}\left(1,1\right)
        + D_C {1\choose 1}\left(1,1\right).
\end{equation}
Using the KMS condition~(\ref{KMS1}) this expression can be
rewritten as,
 \begin{eqnarray}
   D(p) &=& D_R(p) {1\choose 1}\left(1+n(p_0),n(p_0)\right)
          - D_A(p) {n(p_0)\choose 1+n(p_0)}\left(1,1\right).
 \label{Dp}
 \end{eqnarray}

   The fermion propagator $S(P)$, can be obtained by multiplying $D(P)$ with ${\not P}+m$ and replacing
$n(p_0)$ with $-f(p_0)$, the Fermi-Dirac distribution function
\bea
\label{Fermi}
   f(p_0) = {1 \over e^{\beta p_0} +1}\, ,\qquad
   f(-p_0) = \bigl( 1 -f(p_0) \bigr)
   \eea
We have
 \bea
 \label{2x2f}
 S &=&\left(
 \begin{matrix}
    S_{11} & S_{12} \cr
   S_{21} & S_{22}
\end{matrix}\right )
   \nonumber
\eea
with $S_{11} - S_{12} - S_{21} + S_{22} = 0$. The retarded, advanced and correlator components are
\begin{eqnarray} \label{3af}
S_R &=& S_{11} - S_{12} = i\frac{\gamma_4 p_0-i\mathbf{\gamma}\cdot\mathbf{p}+m}{(p_0+i\epsilon)^2-p^2-m^2} \, , \nonumber\\
 S_A &=& S_{11} - S_{21} = i\frac{\gamma_4 p_0-i\mathbf{\gamma}\cdot\mathbf{p}+m}{(p_0-i\epsilon)^2-p^2-m^2} \, , \nonumber\\
 S_C &=& S_{11} + S_{22} = 2\pi(\gamma_4 p_0-i\mathbf{\gamma}\cdot\mathbf{p}+m)(1- 2 f (E_p))\delta(P^2+m^2)\, .
\end{eqnarray}
They satisfy the KMS condition
\bea
S_C(P) &&= (1-2f(p_0))(S_R(P) - S_A(P))
\label{KMS2}
\eea
We can extract the 1PI two-point function, or self energy, by
removing external legs in the usual way. We find,
 \bea
\Pi_R &=& \Pi_{11} + \Pi_{12} \nonumber \\
\Pi_A &=& \Pi_{11} + \Pi_{21} \nonumber \\
\Pi_C &=& \Pi_{11} + \Pi_{22} \label{physPi} \eea where $\Pi_R$ and
$\Pi_A$ are the usual retarded and advanced self energies.
The four CTP
components satisfy the constraint, \bea \Pi_{11} + \Pi_{12} +
\Pi_{21} + \Pi_{22} =0 \nonumber \eea

The eq.(\ref{GammaA}) in CTP formalism takes the form
\begin{equation}
\Gamma[{\cal A}]=\int\frac{d^4Q}{(2\pi)^4}\Big[
-\frac{1}{2}\Pi_{\mu\nu}(Q)^{ab}{\cal A}_\mu^{a*}(Q){\cal A}_\nu^b(Q)+O({\cal A}^3)\Big],
\end{equation}
where the superscripts $a, b$, label the two CTP components. Introducing ${\cal A}=\frac{1}{2}({\cal A}_1-{\cal A}_2)$
and ${\cal A}^\prime={\cal A}_1+{\cal A}_2$, we find that
\begin{equation}
\frac{\delta\Gamma}{\delta{\cal A}_\mu^{\prime*}}\mid_{{\cal A}^\prime=0}
=-\frac{1}{2}(\Pi_{\mu\nu}^{11}+\Pi_{\mu\nu}^{12}-\Pi_{\mu\nu}^{21}-\Pi_{\mu\nu}^{22}){\cal A}_\nu
=-\Pi_{\mu\nu}^R{\cal A}_\nu.
\end{equation}
The recipe to obtain nonlinear responses is given in \cite{kcchou}.











\section{The infrared behavior of the form factor
$C_2(q_1^2,q_2^2,\mathbf{q}_1\cdot\mathbf{q}_2;\omega)$}
\label{IR_formfactor_appendix}

In this appendix, we shall derive the infrared behavior (\ref{infrared}) by calculating the triangle diagrams in Fig.1 with
$\mu=\rho=4$ and $\nu=j$. The trivial color-flavor factor will be suppressed. We shall start with a nonzero Matsubara frequency $\omega=2in\pi T$ and continue it to the upper
edge of the real axis following Baym-Mermin prescription. The amplitude
of the diagram reads
\begin{equation}
\Delta_{4j}(Q_1,Q_2)=-e^2\int\frac{d^3\mathbf{p}}{(2\pi)^3}T\sum_{p_0}
{\rm tr}\gamma_5\gamma_4\left(\frac{1}{{\not P_-}+{\not q_2}}\gamma_j
\frac{1}{{\not P}}\gamma_4\frac{1}{{\not P_-}-{\not q_1}}
+\frac{1}{{\not P_+}+{\not q_1}}\gamma_4\frac{1}{{\not P}}\gamma_i
\frac{1}{{\not P_+}-{\not q_2}}\right),
\label{ampl}
\end{equation}
where $P=(\mathbf{p},i(p_0+\mu))$ and $P_\pm=(\mathbf{p},i(p_0+\mu\pm\omega))$ and
${\not q}=-i\mathbf{\gamma}\cdot\mathbf{q}$.
Since we are only interested the form factor $C_2$ at $\mathbf{q}_1=-\mathbf{q}_2=0$, we expand (\ref{ampl}) according to the powers of spatial momenta $\mathbf{q}_1$ and $\mathbf{q}_2$ and pick up the term proportional to the product of them, i.e.
\begin{equation}
\Delta_{4j}(Q_1,Q_2)=e^2\int\frac{d^3\mathbf{p}}{(2\pi)^3}T\sum_{p_0}
{\rm tr}\gamma_5\gamma_4\left(\frac{1}{{\not P_-}}
{\not q_2}\frac{1}{{\not P_-}}\gamma_j\frac{1}{{\not P}}\gamma_4
\frac{1}{{\not P_-}}{\not q_1}\frac{1}{{\not P_-}}
+\frac{1}{{\not P_+}}{\not q_1}\frac{1}{{\not P_+}}
\gamma_4\frac{1}{{\not P}}\gamma_j\frac{1}{{\not P_+}}
{\not q_2}\frac{1}{{\not P_+}}\right)
\end{equation}
The number of gamma matrices to be traced can be reduced with the aid of the identities
\begin{equation}
\frac{1}{{\not P}_\pm}{\not q}_{1,2}\frac{1}{{\not P}_\pm}
=\frac{-2\mathbf{p}\cdot\mathbf{q}_{1,2}{\not P}_\pm+P_\pm^2{\not q}_{1,2}}{(P_\pm^2)^2}
\end{equation}
$\lbrace\gamma_4,{\not P}_\pm\rbrace=2(p_0+\mu\pm\omega)$ and
$\gamma_4{\not q}_{1,2}=-{\not q}_{1,2}\gamma_4$.
It follows after some algebra that
\begin{equation}
\Delta_{4j}(Q_1,Q_2)=\frac{e^2}{2\pi^2}C_2(0,0,0;\omega)(\mathbf{q}_1\times\mathbf{q}_2)_j
\end{equation}
with
\begin{eqnarray}
C_2(0,0,0;\omega) &=& 8\pi^2\int\frac{d^3\mathbf{p}}{(2\pi)^3}T\sum_{p_0}
\frac{p_0}{P^2}\lbrace\frac{4}{3}p^2\Big[\frac{1}{(P_-^2)^3}-\frac{1}{(P_+^2)^3}\Big]
-\frac{1}{(P_-^2)^2}+\frac{1}{(P_+^2)^2}\rbrace\\\nonumber
&=& 8\pi^2\int\frac{d^3\mathbf{p}}{(2\pi)^3}\oint_C\frac{dz}{2\pi i}f(z)
\frac{z}{z^2-p^2}\lbrace\frac{4}{3}\frac{p^2}{[(z-\omega)^2-p^2]^3}
+\frac{1}{[(z-\omega)^2-p^2]^2}-(\omega\leftrightarrow-\omega)\rbrace,
\label{Cintegral}
\end{eqnarray}
where the angular average of $\mathbf{p}$ has been made and the contour $C$ goes around the imaginary axis clockwisely. The distribution function $f(z)$ is given by (\ref{distr}).
While it is tedious to calculate the residues at the double poles $z=\pm p$ and
the third order poles $z=\pm p\pm\omega$,
there is a short cut to extract the infrared divergent piece as $\omega\to 0$, that is worth elaborating. Mathematically, the contour integral (\ref{Cintegral}) should remain finite
as $\omega\to 0$. The divergent terms of the Laurent expansions in $\omega$ of all residues should cancel each other. On the other hand, the Baym-Mermin
continuation amounts to retain the discrete Matsubara $\omega$ for the residues so that
\begin{equation}
f(\pm p\pm\omega)=f(\pm p) \qquad f^\prime(\pm p\pm\omega)=f^\prime(\pm p)
\qquad ...
\label{bm}
\end{equation}
Then the rational dependence left over is continuated to the real $\omega$-axis without offsetting (\ref{bm}). Consequently, the first few terms of the Taylor expansion of
$f(\pm p\pm\omega)-f(\pm p)$, $f^\prime(\pm p\pm\omega)-f^\prime(\pm p)$ and ... in $\omega$, which are required to cancel the small $\omega$ divergence are missing, resulting in the
infrared divergence after the Baym-Mermin continuation. Such terms are easily identified and we find that
\begin{equation}
C_2(0,0,0;\omega)=\frac{2\pi^2}{3\omega}\int\frac{d^3\mathbf{p}}{(2\pi)^3}
\frac{f^{\prime\prime}(p)-f^{\prime\prime}(-p)}{p}
=\frac{1}{3\omega}\int_{-\infty}^\infty dppf^{\prime\prime}(p)
=\frac{1}{3\omega}
\end{equation}
which gives rise to (\ref{infrared}).

\section{Some technical details behind the one-loop calculation}
\label{oneloop_appendix}

In this appendix, we shall expose some technical details behind the one-loop analysis reported in the section 3.
Substituting eqs.(\ref{AP}) and (\ref{BP}) in, the integrand eq.(\ref{XiA}) may be written as
\begin{eqnarray}
\Xi_{ij}^A(P,Q|m) &=& \frac{I_{ij}}{[(p_0^\prime+\mu)^2-(p'+\mu_5)^2-m^2]
[(p_0+\mu)^2-(p+\mu_5)^2-m^2]}\\\nonumber
&+& \frac{J_{ij}}{[(p_0^\prime+\mu)^2-(p'+\mu_5)^2-m^2]
[(p_0+\mu)^2-(p-\mu_5)^2-m^2]}-(\mu_5\to-\mu_5),
\end{eqnarray}
where $\mathbf{p}'=\mathbf{p}+\mathbf{q}$ and $p_0^\prime=p_0+\omega$,
\begin{eqnarray}
I_{ij} &\equiv& -\frac{1}{4}{\rm tr}\gamma_5[-(p'+\mu_5)\gamma_4+i(p_0^\prime+\mu-m\gamma_4)
\mathbf{\gamma}\cdot\hat{\mathbf{p'}}]\gamma_j
[(p_0+\mu)\gamma_4-i(p+\mu_5)\mathbf{\gamma}\cdot\hat{\mathbf{p}}+m]\gamma_i\\\nonumber
&+& (P'\leftrightarrow P; i\leftrightarrow j)\\\nonumber
&=& i\epsilon_{ijk}(\hat p_k-\hat{p'}_k)
[(p'+\mu_5)(p+\mu_5)+(p_0^\prime+\mu)(p_0+\mu)-m^2]
\end{eqnarray}
and
\begin{eqnarray}
J_{ij} &\equiv& \frac{1}{4}{\rm tr}\gamma_5[-(p'+\mu_5)\gamma_4+i(p_0^\prime+\mu-m\gamma_4)
\mathbf{\gamma}\cdot\hat{\mathbf{p'}}]\gamma_j
[(p_0+\mu)\gamma_4-i(p-\mu_5)\mathbf{\gamma}\cdot\hat{\mathbf{p}}+m]\gamma_i\\\nonumber
&+& (P\leftrightarrow P';i\leftrightarrow j)\\\nonumber
&=& i\epsilon_{ijk}(\hat p_k+\hat{p'}_k)
[(p'+\mu_5)(p-\mu_5)-(p_0^\prime+\mu)(p_0+\mu)+m^2].
\end{eqnarray}
The summation over the Matsubara loop energy $p_0$ is straightforward via a contour integral
and we obtain that:
\begin{eqnarray}
{\cal I}_{ij}(\mathbf{p},\mu_5|m) &\equiv& T\sum_{p_0}\frac{I_{ij}}{[(p_0+\omega+\mu)^2-(p'+\mu_5)^2-m^2]
[(p_0+\mu)^2-(p+\mu_5)^2-m^2]}\\\nonumber
&=& \oint_C\frac{dz}{2\pi i}\frac{I_{ij}}{[(z+\omega+\mu)^2-E_+^{\prime 2}]
[(z+\mu)^2-E_+^2]}\frac{1}{e^{\beta z}+1}\\\nonumber
&=& \frac{i}{2}\epsilon_{ijk}(\hat p_k-\hat{p'}_k)\Big[f(E_+^\prime-\mu)
\frac{(p'+\mu_5)^2+(p'+\mu_5)(p+\mu_5)-\omega E_+^\prime}
{E_+^\prime(E_+^\prime-E_+-\omega)(E_+^\prime+E_+-\omega)}\\\nonumber
&-& f(-E_+^\prime-\mu)
\frac{(p'+\mu_5)^2+(p'+\mu_5)(p+\mu_5)+\omega E_+^\prime}
{E_+^\prime(E_+^\prime-E_++\omega)(E_+^\prime+E_++\omega)}\\\nonumber
&+& f(E_+-\mu)
\frac{(p+\mu_5)^2+(p'+\mu_5)(p+\mu_5)+\omega E_+}
{E_+(E_+-E_+^\prime+\omega)(E_++E_+^\prime+\omega)}\\\nonumber
&-& f(-E_+-\mu)
\frac{(p+\mu_5)^2+(p'+\mu_5)(p+\mu_5)-\omega E_+}
{E_+(E_+-E_+^\prime-\omega)(E_++E_+^\prime-\omega)}\Big],
\label{calI}
\end{eqnarray}
where, the contour $C$ is around the imaginary axis clockwisely. The Baym-Mermin procedure
is employed to continue the imaginary Matsubara energy $\omega=2in\pi T$ to the upper edge of
the real axis to obtain the retarded function. Similarly,
\begin{eqnarray}
{\cal J}_{ij}(\mathbf{p},\mu_5|m) &\equiv& T\sum_{p_0}\frac{J_{ij}}{[(p_0+\omega+\mu)^2-(p'+\mu_5)^2-m^2]
[(p_0+\mu)^2-(p-\mu_5)^2-m^2]}\\\nonumber
&=& \oint_C\frac{dz}{2\pi i}\frac{J_{ij}}{[(z+\omega+\mu)^2-E_+^{\prime 2}]
[(z+\mu)^2-E_-^2]}\frac{1}{e^{\beta z}+1}\\\nonumber
&=& \frac{i}{2}\epsilon_{ijk}(\hat p_k+\hat{p'}_k)\Big[f(E_+^\prime-\mu)
\frac{-(p'+\mu_5)^2+(p'+\mu_5)(p+\mu_5)+\omega E_+^\prime}
{E_+^\prime(E_+^\prime-E_--\omega)(E_+^\prime+E_--\omega)}\\\nonumber
&-& f(-E_+^\prime-\mu)
\frac{-(p'+\mu_5)^2+(p'+\mu_5)(p+\mu_5)-\omega E_+^\prime}
{E_+^\prime(E_+^\prime-E_-+\omega)(E_+^\prime+E_-+\omega)}\\\nonumber
&+& f(E_--\mu)
\frac{-(p-\mu_5)^2+(p'+\mu_5)(p-\mu_5)-\omega E_-}
{E_-(E_--E_+^\prime+\omega)(E_-+E_+^\prime+\omega)}\\\nonumber
&-& f(-E_+-\mu)
\frac{-(p-\mu_5)^2+(p'+\mu_5)(p-\mu_5)+\omega E_+}
{E_-(E_--E_+^\prime-\omega)(E_-+E_+^\prime-\omega)}\Big].
\label{calJ}
\end{eqnarray}
We have then
\begin{eqnarray}
\label{PiA}
\Pi_{ij}^A(Q) &=& e^2\int\frac{d^3\mathbf{p}}{(2\pi)^3}
\lbrace {\cal I}_{ij}(\mathbf{p},\mu_5|m)+{\cal J}_{ij}(\mathbf{p},\mu_5|m)
-{\cal I}_{ij}(\mathbf{p},-\mu_5|m)-{\cal J}_{ij}(\mathbf{p},-\mu_5|m)\\\nonumber
&-& \sum_sC_s[{\cal I}_{ij}(\mathbf{p},\mu_5|M_s)+{\cal J}_{ij}(\mathbf{p},\mu_5|M_s)
-{\cal I}_{ij}(\mathbf{p},-\mu_5|M_s)-{\cal J}_{ij}(\mathbf{p},-\mu_5|M_s)]\rbrace.
\end{eqnarray}
In the static limit, $\omega=0$, the quantity inside the bracket on RHS of (\ref{calI})
reduces to $R/(p-p')$ and that inside the bracket on RHS of (\ref{calJ}) to $R/(p+p')$
with $R$ the same quantity. We have then
\begin{equation}
\frac{\hat{\mathbf{p}}-\hat{\mathbf{p'}}}{p-p'}
+\frac{\hat{\mathbf{p}}+\hat{\mathbf{p'}}}{p+p'}=\frac{\mathbf{q}}{p^{\prime 2}-p^2}.
\label{fraction}
\end{equation}
Upon a shift of the integration momentum, $\mathbf{p}\to\mathbf{p}-\mathbf{q}$
in the terms with $E_\pm^\prime$ inside the Fermi distribution functions, the angular integral becomes elementary.
Bearing in mind that $F_1(q,0)$ is real,
only the principal part of the angular integral is needed and the result is
(\ref{static}) with ${\cal F}(q|m)$ given by (\ref{calF}).

For massless quarks, we may either take the limit of $m\to 0$ of (\ref{PiA}) or compute
${\cal I}_{ij}(\mathbf{p},\mu_5|0)+{\cal J}_{ij}(\mathbf{p},\mu_5|0)$ with the massless
propagator (\ref{propm0}). The contribution from the PV regulators remains intact and the result reads
\begin{eqnarray}
\Pi_{ij}^A(Q) &=& -ie^2\epsilon_{ijk}\int\frac{d^3\mathbf{p}}{(2\pi)^3}
\lbrace\frac{pq_k-\omega p_k}{p}
\Big[\frac{1}{(p-\omega)^2-(\mathbf{p}-\mathbf{q})^2}
     +\frac{1}{(p+\omega)^2-(\mathbf{p}+\mathbf{q})^2}\Big]f(p-\mu+\mu_5)\nonumber\\
&+& \frac{pq_k+\omega p_k}{p}
\Big[\frac{1}{(p+\omega)^2-(\mathbf{p}-\mathbf{q})^2}
     +\frac{1}{(p-\omega)^2-(\mathbf{p}+\mathbf{q})^2}\Big]f(-p+\mu-\mu_5)
     -(\mu_5\leftrightarrow-\mu_5)\rbrace\nonumber\\
&+& \hbox{PV term}
\end{eqnarray}
with ${\rm Im}\omega=0^+$, where we have shifted the integration momentum according to the prescription of the last
paragraph. By symmetry, the angular integral may be performed with $p_k$ replaced by
$\frac{\mathbf{p}\cdot\mathbf{q}}{q^2}q_k$ and we end up with the form (\ref{form}) with
the function $F_1(q,\omega)$ given by (\ref{massless}).


\end{document}